\documentclass[structabstract]{aa}  
\usepackage{graphicx}
\usepackage{txfonts}
\usepackage[]{natbib}
\def\bc{\begin{center}}
\def\ec{\end{center}}

\def\be{\begin{equation}}
\def\ee{\end{equation}}
\def\beq{\begin{eqnarray}}
\def\eeq{\end{eqnarray}}
\def\bfig{\begin{figure}}
\def\efig{\end{figure}}
\def\bnum{\begin{enumerate}}
\def\enum{\end{enumerate}}

\begin{document}
   \title{The properties of cuspy triaxial systems formed due to the 
   radial-orbit instability}


   \author{T. Piffl
          \inst{1}
          \and
          N.Ya. Sotnikova\inst{2}
          }

   \institute{Faculty of Physics, University of Leipzig,
              Linnestrasse 5, 04103 Leipzig, Germany\\
              \email{til@aip.de}
         \and
             Sobolev Astronomical Institute, St. Petersburg State University,
	     Universitetskij pr. 28, 198504 St. Petersburg, Stary Peterhof,
             Russia\\
             \email{nsot@astro.spbu.ru}
             }

   \date{Received ??? ??, 2008; accepted ??? ??, 2008}

 
  \abstract
   {We have investigated the structure and kinematics of triaxial 
   final models formed due to radial-orbit instability from 
   a set of equilibrium anisotropic spherical systems of the 
   Osipkov-Merritt type.}
   {We show that the instability is a natural way to build
    nonspherical systems and thus can be considered as a new technique for
    constructing equilibrium $N$-body models.}
   {The dynamical evolution of the models was followed numerically by means of
   the Dehnen's public code gyrfacON.}
   {We found that the shape of the density profiles for various initial
   spherical models didn't change despite of the systems were drastically 
   rearranged into triaxial configurations and had got the new scale 
   lengths. 
   The cusp was found to be an invariant feature of such a rearranging. Starting 
   from a certain unstable spherical $\gamma$-model we obtain a triaxial 
   system with the same cusp. The end-products are anisotropic at large radii 
   and have anisotropy profiles of the Osipkov-Merrit type. The size of 
   the isotropic core once rescaled corresponds to the value of 
   marginally stable $\gamma$-progenitors and is almost independent of the unstable
   starting point.}
   {We conclude that the end-products reach a new steady state and have 
   quite predictable properties. They can be used as equilibrium models to 
   describe the elliptical galaxies and bulges. They also can be incorporated 
   in $N$-body simulations deal with multicomponent systems.}


   \maketitle
%

\section{Introduction}

Triaxiality is very widespread among stellar systems. For example, there are 
a few perfectly spherical elliptical galaxies. The analysis of their 
morphology and kinematics have shown that ellipticals are at least moderately 
triaxial systems (\citealt{franx_all}). Dark matter haloes formed in 
isolation 
(e.g., \citealt{aguilar_merritt,dubinski_carlberg,katz}) 
or in subregions of larger cosmological simulations 
(e.g., \citealt{frenk_all,warren_all})
through dissipationless collapse of density peaks also demonstrate 
triaxial shapes. The same picture arises in cosmological simulations with 
halo assembling via merging 
(see, e.g., 
\citealt{jing_suto,bailin_steinmetz,shaw_all,allgood_all,bett_all}).

There is one more common feature of real and cosmologically motivated
systems. It is a lack of constant density cores. As a rule such systems have
cuspy density profiles. The stellar surface brigthness of early-type galaxies 
very often continues to rise towards small distances from the centre 
(\citealt{ferrarase_all, lauer_all}). But even a core-like surface brightness 
distribution can be transformed due to projection into a power-law cusp in 
its spatial density (\citealt{lauer_all}). As is known for dark matter haloes
those have universal cuspy density distributions --- the so-called NFW 
profiles with $\rho \propto r^{-\gamma}$ near the centre, where $\gamma=1$ 
(\citealt{nfw96,nfw97}). 

It is very important to have different reliable prescriptions for constructing 
equilibrium models of such systems and to know their structural and dynamical
properties. There are at least two problems which require such models. One
of them deals with photometric and kinematic data fitting. The other
includes many question connected with dynamical evolution of triaxial
stellar systems in a controlled manner and related topics (e.g., the effects
of triaxial halo shape on the disk stability and bar evolution, the survival
of triaxiality during the disk formation and so on). 
The most practical 
models to solve the later tasks are $N$-body realizations of stellar systems. 
There are various and well developped techniques for constructing equilibrium 
models, mainly invented for spherical ones with known DF 
(for the DF construction of models with given density and velocity anisotropy 
profiles see 
\citealt{osipkov,merritt85a,merritt85b,cuddeford,baes_d02,baes_h07}), 
or for the superposition of spherical (or spheroid) and disk components 
(e.g., 
\citealt{hernq93,boily_all,kuijken_dubinski,widrow_dubinski,mcmillan_dehnen,rodionov_as08}).

To build triaxial models with given structural and kinematic parameters is
a far more complicated problem. Jeans theorem and a knowledge of the DF is useful 
only in a few special analytical cases. The most popular numerical technique 
for triaxial system modelling is Schwarzschild's method (\citealt{schwarz}).
Tested initially on core-like systems 
(\citealt{schwarz, statler}) 
it was further applied to obtain self-consistent galaxy models with cusps
(\citealt{merritt_fridman, merritt97}) 
and recently has been generalized to construct models of triaxial
elliptical galaxies embedded in triaxial dark matter haloes 
(\citealt{capuzzo_all}). 
These models based on a library of orbits in a given potential are well
suited to investigate the variety of orbit types presented in a model but
have almost never been used to produce initial conditions for
$N$-body simulations so far.

\cite{holley_all} proposed a technique called
`adiabatic squeezing' to produce cuspy triaxial $N$-body models. They started
with a spherical isotropic Hernquist model (\citealt{hernq90}) and eventually 
were applying a drag to the particle velocities along each principal axis.
During the squeezing, the system retained its initial cuspy density profile
with $\gamma=1$ and displayed a slight radial anisotropy in the velocity
ellipsoid. \cite{rodionov_as08} described a very general prescription to 
construct a triaxial model by means of an iterative method (in original version 
proposed by \citealt{rodionov_s06}) and gave an example of such a model with 
a given density distribution and several constrains on kinematics. 

All these techniques give us equilibrium or near equilibrium triaxial models
but tell nothing about the nature of triaxiality. Triaxiality can arise in a
number of ways. One natural way seen in hierarchical cosmological
simulations is the merger of a number of objects. \cite{moore_all} explored
the generation of triaxial structures formed via merging isotropic
equilibrium spherical haloes of NFW type with varying amounts of angular 
momentum. They investigated the relationship between the value of angular 
momentum and the final shape of configurations (prolate or oblate). Despite
of the final shape their resulting haloes had the same density profile as
the progenitor haloes, independent of their angular momentum. 

Another common way of triaxial system formation is instability. Starting
from equilibrium but unstable conditions a stellar system can
achive a new equilibrium that can be stable. The leading instability pushing
the system towards a new equilibrium and rearranging its initial spherical
density distribution into a triaxial one is the radial-orbit instability
(\citealt{polyachenko_shukhman}; for a physical description of this instability
see \citealt{fridman_polyachenko}). 
It manifests itself in systems with a high degree of velocity anisotropy
(\citealt{merritt_aguilar, meza_zamorano, buyle_all}). This instability is
thought to be the main driver for the formation of elliptical galaxies
(\citealt{vanalbada}). Sufficiently cold initial conditions force system
stars to fall almost radially toward the centre. And if the system has too
many radial orbits such a collapse will result in a triaxial configuration.
Motivated by these ideas \cite{trenti_bertin} fulfilled some numerical
experiments to clarify the following questions: what kind of equilibria can
be achieved in the process of the collisionless collapse starting from
various conditions, what are the structural properties of the 
end-products and are they similar to the real galaxies. 

In this paper we address to initially equilibrium spherical models with cusps 
and different degrees of the velocity anisotropy and analyse shapes, 
structure and kinematics of final triaxial configurations produced via
radial-orbit instability. 

The outline of this paper is as follows. 
In Section 2, we describe the technique for constructing initial equilibrium 
but non-stable cuspy $N$-body models with various degrees of velocity anisotropy 
and the $N$-body code used to study their evolution. 
In Section 3, we investigate the dynamical evolution of our models towards a 
new equilibrium and analyse their structure and kinematics. 
In Section 4, we discuss the properties of the final models in context of 
application to real stellar systems. 
In Section 5, we summarize our main results.


\section{Modelling technique}

\subsection{Family of initial models}

We start with cuspy spherical systems from the one-parameter family
$\gamma$-models (\citealt{dehnen93,tremaine_all94}).
These models follow a density profile given by 

\be
\label{eq_gamma}
\rho(r) = \frac{(3-\gamma)a}{4\pi}\frac{M}{r^\gamma(r + a)^{4-\gamma}}\, ,
\ee
where $M$ --- the total model mass and $a$ --- the typical scale length. The
density of the models behaves as $\rho \propto r^{-4}$ for large radii and as 
$\rho \propto r^{-\gamma}$ near the centre, thus $\gamma$ characterizes how
a central cusp is strong.

We have limited our study to models with $\gamma=1$ (`weak-cusp') and
$\gamma=3/2$ ('moderate-cusp'). The former model is the so-called Hernquist
sphere (\citealt{hernq90}). Surface density of both models is shown 
(\citealt{dehnen93}) to agree quite well with the de Vaucoulers profile 
(\citealt{devaucouleurs}) that describes the surface brightness of many
ellipticals and bulges of spirals. 

We considered a set of anisotropic models of the Osipkov-Merritt type
(\citealt{osipkov,merritt85a,merritt85b}). 
In this case the anisotropy parameter $\displaystyle 
\beta = 1 - \sigma_{\mathrm{t}}^2/ 2\sigma_{\mathrm{r}}^2$, 
where $\sigma_{\mathrm{t}}^2$ and $\sigma_{\mathrm{r}}^2$ are the tangential
and radial components of the velocity dispersion, is a function of radius in
the form (\citealt{osipkov,merritt85a,merritt85b})
\be
\label{eq_om}
\beta_{\rm OM} = \frac{r^2}{r^2 + r_{\rm a}^2} \, .
\ee
The anisotropy radius $r_{\rm a}$ cut off the isotropic core from the
radial-orbit dominated region outside $r_{\rm a}$.

Models with a given density distribution have two critical values for 
$r_{\rm a}$. One value seperates physical models with a strictly positive 
DF from unphysical ones. Another value marks the stability boundary.
\cite{merritt_aguilar, meza_zamorano, buyle_all} set
numerically the stability threshold for various models including
$\gamma$-models of Osipkov-Merritt type. For $\gamma = 1$ the critical value
of $r_{\rm a}$ is $\approx 1.1$, for $\gamma = 3/2$ the stability threshold was
found to be  at $r_{\rm a} \approx 0.8$ (\citealt{meza_zamorano}). We consider 
long time dynamical evolution of physical models lying below the stability 
threshold. 

In all numerical simulations we used the following dimensionless units: 
$G = 1$, $M = 1$, $a = 1$. These units can be converted in physical ones  
through typical observational parameters of elliptical galaxies --- mass and
effective radius that characterizes the de Vaucoulers profile. For
$\gamma=1$ effective radius $r_{\rm e} = 1.815 a$ (\citealt{hernq90}), for
$\gamma=3/2$ it can be derived from the cumulative surface density
(\citealt{dehnen93}) $r_{\rm e} = 1.276 a$. 
If we adopt $M = 2 \cdot 10^{11} M_{\sun}$, $r_{\rm e}=3$~kpc as typical 
parameters for bright ellipticals it will give us the scale unit  
$r_u = a = 1.65$~kpc, the time unit $t_u = 2.24$~Myr, 
the velocity unit $v_u = 721.3$~km/s \,for $\gamma = 1$, 
and 
$r_u = a = 2.35$~kpc, $t_u = 3.80$~Myr, $v_u = 604.7$~km/s 
\,for $\gamma=3/2$. 

\cite{meza_zamorano, buyle_all} expressed the evolution of
their models in dimensionless units of half-mass dynamical time $T_{\rm h}$.
It is evaluated at the half-mass radius $r_{1/2}$ (\citealt{binney_tremaine}).
For $\gamma=1$ it is equal $T_{\rm h} \approx 8.3$ 
and $T_{\rm h} \approx 4.9$ for $\gamma=3/2$. \cite{dehnen93} noticed that
the ratio $r_{\rm e} / r_{1/2} \approx 0.75$ is practically the same for all
$\gamma$-models. This means that if we fix the physical parameters (mass and
effective radius) we will obtain the same physical dynamical time scale for
all models. For $\gamma = 1$ with adopted parameters it will be $T_{\rm h}=
18.7$~Myr and $T_{\rm h} = 18.8$~Myr for $\gamma = 3/2$.

We followed the dynamical evolution of the models further than in
\cite{meza_zamorano, buyle_all}, for some models  -- up to 
$t \approx 1000-1400$ (i.e. $\approx (100-200) T_{\rm h}$). It concerns the 
models just below the stability threshold. They were approaching the new 
steady state very slowly exhibiting a downfall to a new constant level in 
the behaviour of the axis ratio only after $70-120$ dynamical times.

\subsection{Initial data sets}

An $N$-body realization of an initialy spherically symmetric system was
created in a conventianal way by using the DF. For Osipkov-Merritt models
the DF has a form
\be
f({\cal E},\,L) = f(Q) \, ,
\label{DF_OM}
\ee
where $Q = {\cal E} - L^2/2r_{\rm a}^2$, 
${\cal E} = \Psi - \frac{1}{2} v^2$, $\Psi$ is the negative of the
gravitational potential, and $L=rv_{\rm t}$ is the angular momentum with
$v_{\rm t}$ the tangential velocity components. The DF in the 
form~(\ref{DF_OM}) for a given density profile can be found through the 
technique (\citealt{osipkov,merritt85a,merritt85b}) which is an extension 
of Eddington's \citeyearpar{ed1916} inversion technique to obtain the DF 
for an isotropic spherical system.

The initial conditions in the $N$-body problem suggest specifying the mass,
position in space, and three velocity components 
($v_r$, $v_{\varphi}$, $v_{\theta}$) for each particle. 

The mass of all particles was assumed to be the same.
As the DF is specified the particle positions and velocities are initialized
by sampling the DF. The particle coordinates are naturally determined in 
accordance with the density profile~(\ref{eq_gamma}). The usual way to do 
this is to invert the cumulative mass profile for the radius.

For velocity components it is enough to sample $v_r$ and 
$v_{\rm t} = \sqrt{v_{\varphi}^2 + v_{\theta}^2}$ and to choose the random
direction for $v_{\rm t}$ in the plane which is perpendicular to the
radius-vector. The algorithm for velocity sampling in anisotropic systems of
more general type than the Osipkov-Merritt models is
breifly described in \cite{mcmillan_dehnen} and implemented in the public code
{\tt mkhalo} which we have taken from the Walter Dehnen's part of the public NEMO 
package (http://astro.udm.edu/nemo; \citealt{teuben}).

\subsection{Computational method}

We investigated the dynamical evolution of our models by using the $N$-body 
code {\tt gyrfalcON} (\citealt{dehnen00,dehnen02}) that combines a 
hierarchical tree method (\citealt{barnes_hut}) and a fast multipole method 
(\citealt{greengard_r}). This code has complexity ${\cal O}(N)$ and is faster 
than a standard tree code by at least a factor 10. Its implementation was 
taken from the public NEMO package 
(\citealt{teuben}).

For almost all models we chose the number of particles $N = 300\,000$.
The tolerance parameter $\theta$ that is responsible for the accuracy
of calculating the gravitational force was fixed at $0.6$ in all our
simulations. Calculations were performed in a single-time-step mode. In this
case momentum is exactly conserved. The time step and softening length were 
taken $dt = 1 / 2^6$ and $\epsilon = 0.04$. The choice of these parameters 
was in agreement with the recommendations of \cite{rodionov_s05}.

\section{Properties of final equilibrium models}

We performed the simulations on a long time scale (up to about 200 dynamical
time for some models), and used the temporal behaviour of the axis ratios 
$b/a$ and $c/a$ as indicators of new steady state settling. Till now such
investigations were concentrated on finding numerically the stability
threshold of spherical anisotropic models 
(\citealt{merritt_aguilar,dejonghe_merritt,meza_zamorano,meza02,buyle_all}). 
Triaxial end-products formed due to the radial-orbit instability from 
initially equilibrium systems were commonly considered as a `garbage'. 
Below we studied the structure and kinematics of such systems settled in 
a new equilibrium in details and determined the relationship between
properties of initial and final models.

\subsection{Model shapes}

To determine the shape of the particle distributions we
used a technique descibed by \cite{dubinski_carlberg} and \cite{katz} (see
also \citealt{meza_zamorano,buyle_all}). First of all we transformed the
input $N$-body snapshot at specified time moments to a coordinate system which
diagonalizes the moment of inertia tensor 
$I_{ij} = \sum{x_ix_j}/r^2$ of a specified subset of particles 
within a certain sphere. For the sphere radius we chose $r = 5$. The
centroid of the subset was translated to the origin. After diagonalization 
the principal components of the inertia tensor  
$I_{xx} \geq I_{yy} \geq I_{zz}$ 
were used to
calculate the preliminary values of the axis ratios
\be
\frac{b}{a} = \left(\frac{I_{yy}}{I_{xx}}\right)^{1/2} \,\,\,\,
\mathrm{and} \,\,\,\,\,\,\,\,
\frac{c}{a} = \left(\frac{I_{zz}}{I_{xx}}\right)^{1/2} \, .
\ee
After that we determined the ellipsoidal radius $q$ as 
$q = (x^2 + y^2/(b/a)^2 + z^2/(c/a)^2)^{1/2}$ and calculated the eigenvalues
and eigenvectors of inertia tensor for a subset of particles with $q \leq
5$. Tensor components were found as $I_{ij} = \sum(x_i x_j/q^2)$. 
The new axis ratios were used as the conditions for the next iteration.
We repeated this procedure until the error of both axial ratio
determination became less than $10^{-4}$.

\begin{figure*}[th!]
   \centering
   \includegraphics[angle=-90,width=\textwidth]{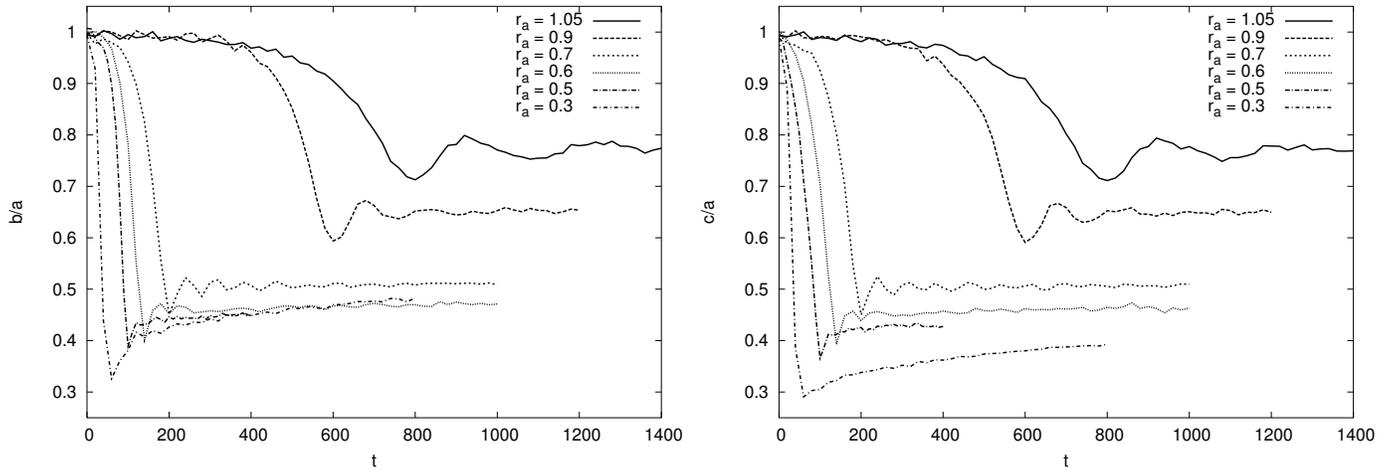}
      \caption{Longtime evolution of the intermediate $b$ and minor $c$ 
               to major $a$ axis ratios for the $\gamma=1$ models with 
	       different values of initial anisotropy radius $r_{\mathrm a}$. 
	       The axes lengths are iteratively calculated from the 
	       ellipsoidal density distribution using the moment of 
	       inertia tensor.
              }
         \label{bc_a1.0}
   \end{figure*}

\begin{figure}[t]
   \centering
   \includegraphics[angle=-90,width=9cm]{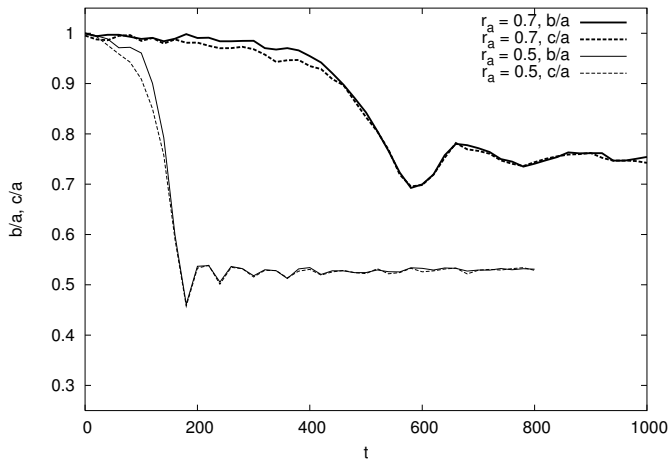}
      \caption{Longtime evolution of the $b/a$ and $c/a$ 
               ratios for the $\gamma=3/2$ models with two 
	       values of initial anisotropy radius $r_{\mathrm a}$. 
	       The axes length are iteratively calculated from the 
	       ellipsoidal density distribution using the moment of 
	       inertia tensor.
              }
         \label{bc_a1.5}
   \end{figure}

Figures~\ref{bc_a1.0}~and~\ref{bc_a1.5} demonstrate the temporal behavior of
the ratios between the intermediate and major axes and minor and major axes
for two $\gamma$-models with different values of the anisotropy radius 
$r_{\rm a}$. 

Reaching the constant value of the axes ratio was considered 
as a new equilibrium. Some models with the anisotropy radius just below the
stability threshold were approaching this state on a time scale longer than
$T_{\rm h} = 50$. As a result we obtained more extended end-products. For
example, for $\gamma=1$ the stability threshold was set as $r_{\rm a}=1.1$ 
(\citealt{meza_zamorano,buyle_all}). Starting from $r_{\rm a}=1.05$ we have 
got after 150 dynamical times ($t \approx 1200$) a perfectly prolate model 
with $b/a,\, c/a \approx 0.77$. The model with the anisotropy radius far 
from the stability threshold ($r_{\rm a}$) seems to be in nonequilibrium
even after 100 dynamical times. A slow axes evolution is not noticeble on a
short time scale ($T_{\rm h}=50$) but is clearly seen during a longer period.

%

%

In the Table~\ref{table:1} we summarize the properties of all sets of
simulations. The values of $b/a$ and $c/a$ are averaged over last 100 units
of the time. They are very close to each other. It means that  
all the end-products of the initially unstable $\gamma$-models have an 
extremly prolate shape with a slight triaxiality for 
the model with $\gamma=1$ and $r_{\rm a}=0.5$ and for the nonequilibrium 
model with $\gamma=1$ and $r_{\rm a}=0.3$.

%
\begin{table}[ht!]
\caption{Properties of the final models}
\label{table:1}
\centering
\begin{tabular}{r l r c c c c c c}
\hline\hline
$\gamma$ & $r_{\rm a}$ & $T_{\rm end}$ & $r_{\rm a,\,f}$ & $\delta$ & $a$ 
& $b/c$ & $c/a$ & $r_{\rm a,\,f}/a$ \\
\hline
1   & 1.05 & 1400 & 1.46 & 2.11 & 1.11 & 0.78 & 0.77 & 1.31 \\
1   & 0.9  & 1200 & 1.59 & 2.07 & 1.24 & 0.65 & 0.65 & 1.29 \\
1   & 0.7  & 1000 & 1.90 & 1.94 & 1.53 & 0.51 & 0.51 & 1.24 \\
1   & 0.6  & 1000 & 2.00 & 1.79 & 1.83 & 0.47 & 0.46 & 1.10 \\
1   & 0.5  & 400  & 2.08 & 1.79 & 1.93 & 0.45 & 0.43 & 1.07 \\
1   & 0.3  & 800  & 1.91 & 1.64 & 2.03 & 0.48 & 0.39 & 0.94 \\
3/2 & 0.7  & 1000 & 1.21 & 2.02 & 1.06 & 0.75 & 0.75 & 0.92 \\
3/2 & 0.5  & 1000 & 0.97 & 2.09 & 1.44 & 0.53 & 0.53 & 0.84 \\
\hline
\end{tabular}
\end{table}

\subsection{Density profiles}

The most remarkable feature of the end-products is their density profiles.
During reaching a new equilibrium the $\gamma$-models retain their initial
density profiles. Once the new scale length $a$ is found all density profiles 
of the final models can be described by a law likes~(\ref{eq_gamma})
but modified in the case of triaxiality as 

\be
\label{eq_gamma_tr}
\rho(q) = \frac{(3-\gamma)a^3}{4\pi bc}\frac{1}{q^\gamma(q + a)^{4-\gamma}}\, ,
\ee
where $q$ is ellipsoidal radius.

\begin{figure*}[ht]
   \centering
   \includegraphics[angle=-90,width=\textwidth]{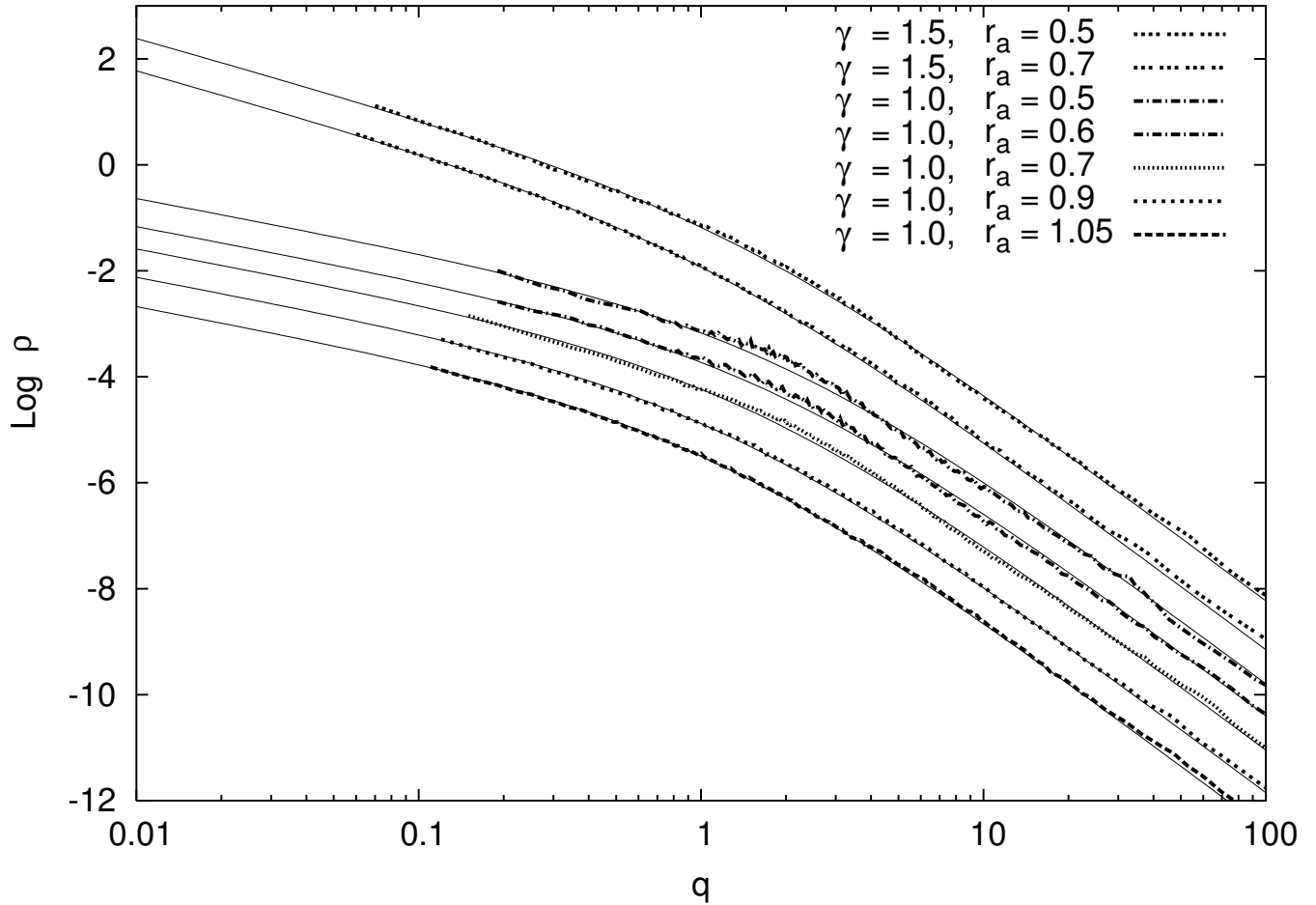}
      \caption{Density of the models plotted as a function of ellipsoidal
               radius $q$. The curves show the density profiles for the final
               triaxial models formed via the radial-orbit instability 
	       from the initial $\gamma$-models with different values of
	       anisotropy radius $r_{\mathrm a}$. The thin solid lines show 
	       the fitting of profiles by formula~(\ref{eq_gamma_tr}).
	       Curves are shifted from one another by dex=0.5 
	       (from top to botom) for convenience.
              }
         \label{dens_fit}
   \end{figure*}

Figure~\ref{dens_fit} shows the density of the end-products as a function 
of an ellipsoidal radius $q$. We demonstrate also the fitting of 
some final density profiles by formula~(\ref{eq_gamma_tr}) 
with $b/a$ and $c/a$ taken at the end of simulations. The profiles can be 
fitted with an accuracy of better than 1\%. The models exhibit an
unchanging density profiles over $\sim 3-4$ orders of magnitude in radius 
after settling to the new equilibrium, and they do not evolve away from the 
original cusps. The density slope seems to be an invariant feature of these 
models despite of their drastical rearranging into the triaxial 
configurations. Starting from a spherical model with a certain $\gamma$ 
cusp we obtain a triaxial model with the cusp of the same shape.

%

\subsection{Kinematics and velocity anisotropy}

\begin{figure*}
   \centering
   \includegraphics[angle=-90,width=\textwidth]
                   {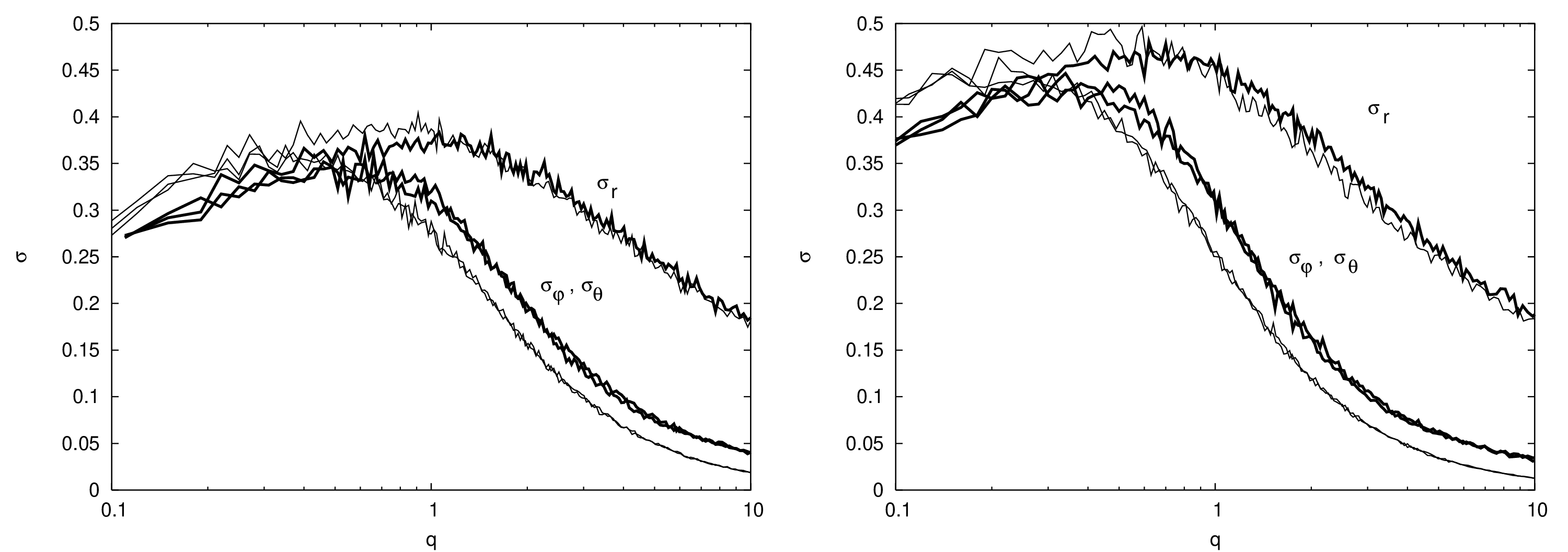}
      \caption{The dependance of three moments of the velocity distribution,
               namely $\sigma_r$, $\sigma_{\varphi}$ and $\sigma_{\theta}$,
               as a function of ellipsoidal radius $q$. Thin curves ---
               the initial profiles, thick curves --- the profiles for 
	       end-products. Left --- $\gamma=1$, $r_{\rm a} = 1.05$; 
	       right --- $\gamma=3/2$, $r_{\rm a} = 0.7$.
              }
         \label{disp}
   \end{figure*}

The velocity dispersions  
($\sigma_r$, $\sigma_{\varphi}$, $\sigma_{\theta}$) as a function of radius
in Osipkov-Merritt models can be easy computed as the corresponding moments
of the DF in the form $f(Q)$ (\citealt{merritt85a,merritt85b}). The 
$\gamma$-models of the Osipkov-Merrit type have a distinctive hump in their
velocity dispersion profiles provided $\gamma < 2$. The same feature
manifests itself in the radial profiles of the projected velocity dispersion
(\citealt{carollo_all}). 
The end-products exhibit the same hump in their velocity dispersion
profiles like their spherical progenitors (Fig.~\ref{disp}). Despite of their 
nonspherical shape all models demonstrate the equivalency between 
$v_{\varphi}$ and $v_{\theta}$ velocity components as the spherical models 
with DF in the form $f({\cal E},\,L)$.

%

\begin{figure*}
   \centering
   \includegraphics[angle=-90,width=\textwidth]
                   {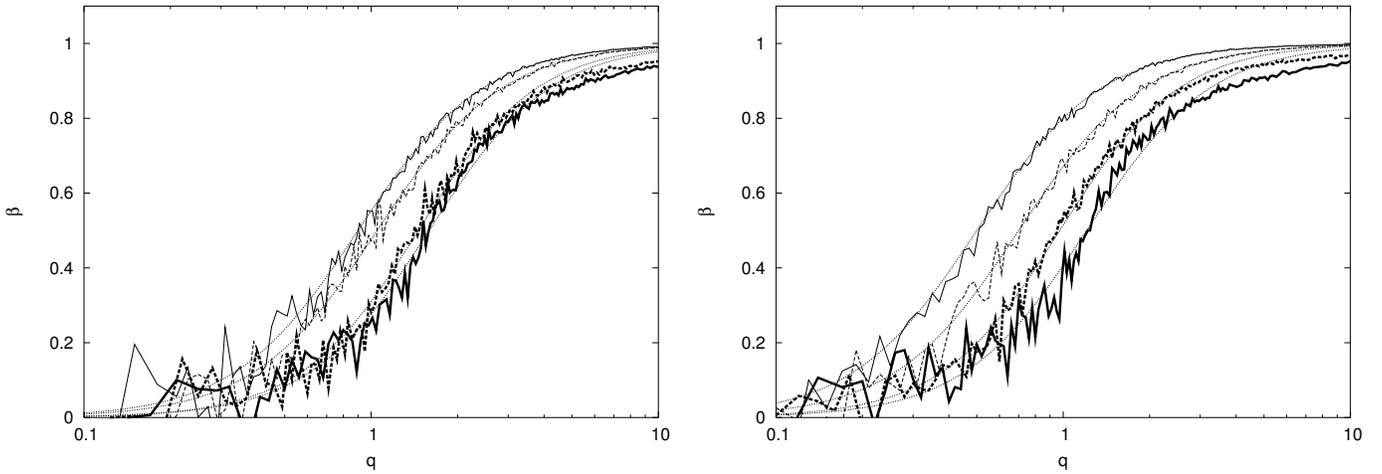}
      \caption{The velocity anisotropy parameter as a function of
               ellipsoidal radius $q$. Thin curves ---
               the initial profiles, thick curves --- the profiles for 
	       end-products, thin dotted lines show profiles obtained by
               fitting the data by formula (\ref{eq_om-mod}). 
	       Left --- $\gamma=1$; solid lines --- $r_{\rm a} = 0.9$,
               dashed lines $r_{\rm a} = 1.05$. Right --- $\gamma=3/2$, 
	       solid lines --- $r_{\rm a} = 0.5$, 
	       dashed lines $r_{\rm a} = 0.7$.
              }
         \label{beta}
   \end{figure*}

All final models display a radial anisotropy in the velocity ellipsoid.
We choose to represent the anisotropy profiles with more general law than
(\ref{eq_om})
\be
\label{eq_om-mod}
\beta = \frac{q^\delta}{q^\delta + r_{\rm a,\,f}^\delta} \, ,
\ee
with $r_{\rm a,\,f}$ and $\delta$ being free parameters. The results of
fitting are summarized in the Table~\ref{table:1}. All models obtained from
progenitors laying just below the stability threshold have the anisotropy
profile with $\delta\approx 2$, e.g. they are of Osipkov-Merritt type.
Fig~\ref{beta} shows some examples. The profiles were fitted over the range
$q < 5$. The deviation from (\ref{eq_om-mod}) are observed only at large 
distances $q > 5$ from the center and are due to a slight predominance of 
tangential orbits there as compared with Osipkov-Merritt models.

%

\section{Discussion}

(1) Triaxial models formed due to instabilities from spherical anisotropic
equilibrium systems were commonly thought as a `garbage'. Their structure and 
kinematics weren't investigated in details. We showed that such models can
be considered as equilibrium ones and used for studying their dynamical
evolution and related topics concerned with dynamics of multicomponent
systems.

(2) All our models have a near prolate shape without any strong triaxiality.
Such a shape is a common feature of cosmologically motivated dark halos. 
There have been many theoretical papers published over the years which
examined the subject of halo shapes. Most authors found that halos tend to
be prolate. For example, using six high resolution dissipationless
simulations with a varying box size in a flat LCDM universe
\cite{allgood_all} found that most halos in their simulations were prolate in
shape with very few oblate halos. But our prolate models, that are based on 
initial spherically symmetric $\gamma$-models, are rather suit to describing 
the density distribution of ellipticals and bulges. 

Several authors discussed the problem of the intrinsic shape of elliptical 
galaxies (e.g., \citealt{franx_all,vincent_ryden,padilla_strauss}).
Many of them agree that the observed distribution of ellipticities can be 
reproduced by suggesting the triaxiality of these objects. It is concluded 
that any acceptable distribution is dominated by nearly-oblate spheroidal 
rather than nearly-prolate spheroidal systems. \cite{padilla_strauss} have 
recently reexamined the underlying shapes of elliptical galaxies in the 
SDSS Data Release 6 from the observed distribution of projected galaxy
shapes, taking into account the effects of dust extinction and reddening. 
The elliptical galaxy data were found to be consistent with oblate spheroids, 
with a correlation between luminosity and ellipticity: more luminous 
ellipticals tend to be rounder, although ellipticals are oblate at all 
luminosities. This result is in agreement with an analysis by
\cite{vincent_ryden} based on SDSS Data Release 3 although later authors 
found that the fainter de Vaucoulers galaxies are best fitted with prolate 
spheroids with mean axis ratio $\langle (b,c) / a \rangle \approx 0.51$. 

\cite{mendez_all} performed two-dimensional photometric decomposition 
of the galaxy surface brightness distribution to derive the structural 
parameters of disks and bulges for 148 unbarred S0-Sb galaxies. They obtained 
the probability distribution function of the intrinsic equatorial ellipticity 
of bulges and found that about 80\% of bulges in their sample are not oblate 
but triaxial spheroids. This is consistent with several previous findings.

Thus, in the context of observational data our models can be related only
with the lowest luminosity ellipticals.

(3) All our final models are strongly elongated with 
$\langle (b,c) / a \rangle \approx 0.5-0.8$. It's true even for the 
models starting from the conditions just below the level of stability. 
For $r_a = 1.05$ we folowed the dynamical evolution of the system further 
($T_{\rm end} \approx 170 T_h$) than it was done by \cite{meza_zamorano}
and \cite{buyle_all} and obtained a very elongated end-product. 

(4) One of the main points of our investigation is that the shape of the 
density profile didn't change. The end-product structure can be predicted in
a quite definite way and depends only on initial model properties.
\cite{moore_all} noticed that their triaxial remants formed via
merging keep central density slopes of progenitors. 
\cite{boylan_ma} claimed that if the cuspy halo had started out with an NFW 
form with $\gamma=1$, the merged halo also had a form close to NFW one, with a
slightly shallow inner cusp of $\rho \propto r^{-0.7}$ instead of $r^{-1}$. 
\cite{holley_all}, who had used a squeezing technique for constucting 
triaxial systems, also concluded that their models retained the given 
$\gamma$ character after squeezing had terminated. It seems the cusp shape 
is an invariant feature of any matter rearranging as though the particles
retain the memory of the progenitor density profile.

(5) The dark halos obtained in cosmological simulations have a small degree 
of anisotropy far from the center 
(\citealt{cole_lacey,colin_all,fukushige_makino,diemand_all}). 
We know a little about 3D velocity distributions in ellipticals.
But modern methods for modelling of observational data allow to reconstruct
such distributions. Combining surface brightness photometry and long-slit 
absorbtion line kinematic data, \cite{delorenzi_all} have recently presented 
a dynamical study of NGC 4697, intermediate-luminosity, E4 galaxy. Their best 
fitting models were only slightly radially anisotropic, with the anisotropy 
parameter increasing to $\beta \simeq 0.5$ at large radii.

The kinematic properties of the prolate models under study aren't similar to 
the properties of the cosmological dark halos and ellipticals. The possible
explanation is that we started out from very anisotropic models of the OM 
type. The velocity anisotropy profile on the periphery was produced by purely 
radial orbits ($\beta$ tends to 1). The end-products are also anisotropic at 
large radii but $\beta$ is less than 1 (see Fig~\ref{beta}). However, as one 
can see from the Table~\ref{table:1}, the size of the isotropic core 
increased for all models and once rescaled it corresponds to the value for 
marginally stable $\gamma$-progenitors and almost independant on a starting 
unstable point. As opposed to this picture, in a merging scenario the particle 
velocities, being initially isotropic, remain mostly isotropic near the 
centre of the merger remnant, but become mildly anisotropic ($\beta < 0.4$) 
with increasing radius due to radial infalls (\citealt{boylan_ma}).

Velocity dispersion profiles of our models demonstrate one remarkabe feature
--- the equivalency between $v_{\varphi}$ and $v_{\theta}$ velocity
components. This fact in combination with a model shape may help 
to construct the DF of such prolate systems in the form $f({\cal E},\,L)$.

(6) We followed the evolution of our models further than in previous
investigations (\citealt{meza_zamorano,buyle_all}). These authors studied
the stability of their models by using an $N$-body code based on the
`self-consistent field' method (\citealt{hernquist_ostriker}), in which the
density and the gravitational potential are expanded in a biorthogonal set
of some basis functions. The using of spherical harmonics provided a
reflection symmetry. 
We used a tree-like code (\citealt{teuben,dehnen00,dehnen02}) and reached a 
definitive equilibrium only by a special symmetrization procedure that 
preserves a reflection symmetry of a model during a simulation 
(the manipulator {\tt symmetrize} in Dehnen's {\tt gyrfalcON}). One have to 
keep this fact in mind when modelling such systems.

   \begin{figure}[!ht]
   \centering
   \includegraphics[angle=-90,width=9cm]{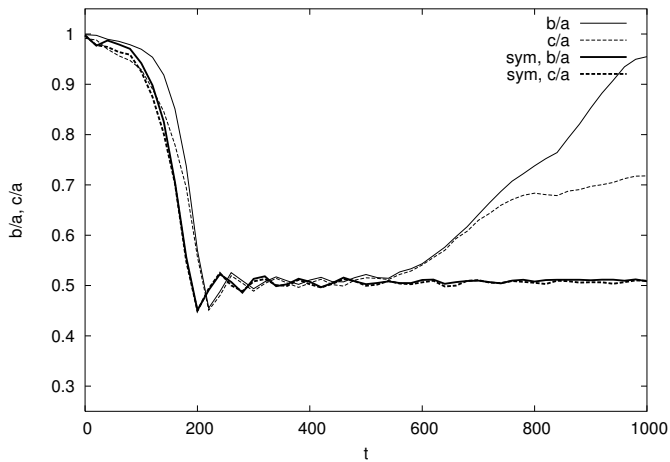}
      \caption{Longtime evolution of the $c/a$ ratio for the $\gamma=1$ 
               models with $r_{\mathrm a}=0.7$. Thin curves --- calculations
               without symmetrization, thick curves --- calculations with 
	       {\tt symmetrize} manipulator.
	       The axes length are iteratively calculated from the 
	       ellipsoidal density distribution using the moment of 
	       inertia tensor.
              }
         \label{sym_nosym}
   \end{figure}
%

(7) \cite{merritt_fridman,merritt97} considered the question about the degree
to which triaxiality can be supported in galaxies with high central
densities. They concluded that triaxiality is inconsistent with a strong
cusp due to rapid chaotic mixing and a real galaxy might choose axisymmetry
with a tendency to be nearly prolate spheroidal, nearly prolate, or nearly
spherical.
All our models formed via the radial orbit instability are axissymmetric with
a prolate shape. This form is reached on a time scale of about $\sim
50\,T_{\rm h}$. After that it can be preserved only by a `hand-made'
maintenance of a reflection symmetry. Without this maintenance some of our 
systems undergo the instability and start evolving towards nearly oblate
spheroids (see~Figure~\ref{sym_nosym}). It is known that instability criteria
are the strongest constraints on structural and kinematic parameters of
stellar systems and instability is a natural way for equilibrium stellar
system formation. That is why the models constructed here need further
investigations. Futher studies will analyse the long time orbital evolution 
of particles to study orbit families consistuting these models and the
degree of chaos.

\section{Summary}

Our results can be summarized as follows:

\begin{enumerate}
\item The radial orbit instability is a natural way to build
nonspherical systems and can be considered as a new technique for
constructing equilibrium $N$-body models.
\item This models have quite predictable properties because the shape of 
the density profiles doesn't change in the process of instability. For
example, the initial cusp survives.  
\item New steady state models have a definitely prolate shape.
\item The degree of anisotropy is diminishing and end-products are more
isotropic.
\end{enumerate}

\begin{acknowledgements}
      This work was partially supported by the Russian Foundation for Basic
      Research (grant 06--02--16459) and by grant from President of the 
      Russian Federation for support of Leading Scientific Schools 
      (grant NSh--8542.2006.02).
\end{acknowledgements}

\end{document}